\documentclass[aps,pre,twocolumn,superscriptaddress,showpacs]{revtex4-1}
\usepackage{graphicx,amsmath,amssymb}
\renewcommand{\vec}[1]{\mathbf #1}
\newcommand{\dd}{\mathrm d}
\begin{document}

% Use the \preprint command to place your local institutional report
% number in the upper righthand corner of the title page in preprint mode.
% Multiple \preprint commands are allowed.
% Use the 'preprintnumbers' class option to override journal defaults
% to display numbers if necessary
%\preprint{}

%Title of paper
\title{Tube width fluctuations of entangled stiff polymers}

% repeat the \author .. \affiliation  etc. as needed
% \email, \thanks, \homepage, \altaffiliation all apply to the current
% author. Explanatory text should go in the []'s, actual e-mail
% address or url should go in the {}'s for \email and \homepage.
% Please use the appropriate macro foreach each type of information

% \affiliation command applies to all authors since the last
% \affiliation command. The \affiliation command should follow the
% other information
% \affiliation can be followed by \email, \homepage, \thanks as well.
\author{Jens Glaser}
 \email{jglaser@umn.edu}
 \altaddress{present
  address: CEMS, University of Minnesota, 421 Washington Avenue SE,
  Minneapolis MN 55455}
 \author{Klaus Kroy}
 \affiliation{Institut
  f\"ur Theoretische Physik, Universit\"at Leipzig, PF 100920, 04009
  Leipzig, Germany}

%Collaboration name if desired (requires use of superscriptaddress
%option in \documentclass). \noaffiliation is required (may also be
%used with the \author command).
%\collaboration can be followed by \email, \homepage, \thanks as well.
%\collaboration{}
%\noaffiliation

\date{\today}

\begin{abstract}
  The tube-like cages of stiff polymers in entangled solutions have
  been shown to exhibit characteristic spatial heterogeneities. We
  explain these observations by a systematic theory generalizing
  previous work by D. Morse (Phys. Rev. E 63:031502, 2001).  With a
  local version of the binary collision approximation (BCA), the
  distribution of confinement strengths is calculated, and the
  magnitude and the distribution function of tube radius fluctuations
  are predicted. Our main result is a unique scaling function for the
  tube radius distribution, in good agreement with experimental and
  simulation data.
\end{abstract}

\maketitle

\section{Introduction}
Entangled solutions of stiff polymers are minimal model systems to
generate a fundamental understanding of the origin of the mechanical
properties of the cytoskeleton. This complex polymer scaffold
maintains the stability and integrity of animal cells and is comprised
of three types of semiflexible filaments, microtubules, actin, and
intermediate filaments, with backbone diameters in the nanometer range
but persistence lengths on the order of $10^{-1}\dots 10^{3} \mu m$
\cite{Gittes1993,Isambert1995,Schopferer2009}.  Single stiff
biopolymers exhibit a rich mechanical response
\cite{Ghosh2007,Emanuel2007,Hallatschek2007a,Obermayer2009}.  In-vitro
reconstituted solutions of such biopolymers hint at how cells can
acquire a considerable macroscopic strength from a purely topological
microscopic constraint and thermal fluctuations, utilizing a minimum
amount of material. Though the individual polymers only have to
respect a simple constraint, namely the mutual impenetrability of the
polymer backbones, complex soft-solid mechanical behavior arises at
densities that would correspond to a very dilute gas without
polymerization and a certain flexibility allowing for thermal backbone
undulations.  To deform an entangled polymer, surrounding polymers
need to be pushed out of the way, as familiar from knotted strings.
This mechanism leads to confinement of the individual polymers in
effective tube-like cages, from which they only escape very slowly by
a snake-like motion called reptation
\cite{Edwards1967,DeGennes1971}. The suppression of chain motion
perpendicular to the tube backbone is responsible for the remarkable
integrity of the transient network.  A microscopic derivation of this
confinement poses formidable theoretical challenges, and there has so
far been little progress beyond the introduction of basic topological
invariants characterizing polymer entanglement
\cite{Edwards1967a,Muller-Nedebock1999} and a phenomenological
primitive path analysis \cite{Everaers2004}.

\begin{figure}[t!]
\centering\includegraphics[width=\columnwidth]{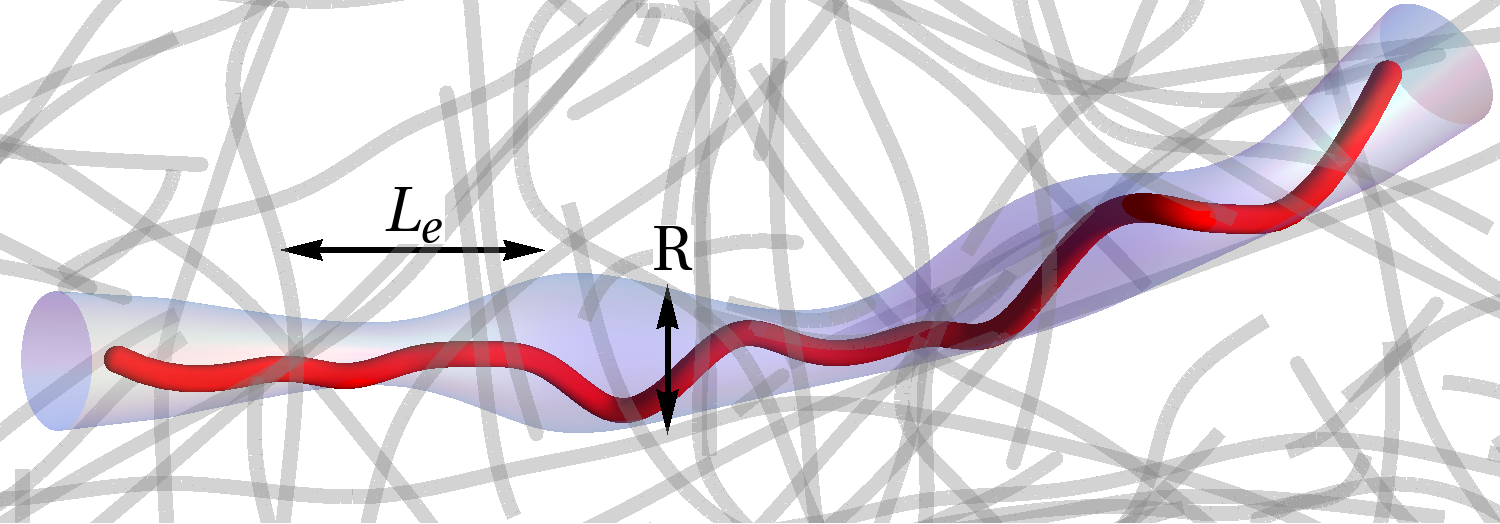}
\caption{(Color online) Test polymer in a background solution,
  confined into a tube of spatially varying radius $R(s)$. The
  chemical distance $L_e$ indicates the characteristic scale of the
  tube heterogeneities.}
\label{fig:tube}
\end{figure}

Nevertheless, self-consistent approximations for the dynamics of rigid
\cite{Sussman2011, Sussman2011a} and the equilibrium statistical
mechanics of semiflexible \cite{Morse2001} topologically entangled
chains have been worked out, and these treatments can predict salient
properties of the reptation dynamics and the postulated tube. For
stiff but not rigid polymers, with a large but finite persistence
length $l_p$, the tube confinement of the transverse fluctuations of a
representative test chain is implemented by a harmonic potential of
stiffness $\phi$. The confinement geometry is characterized by an
entanglement (or collision) length $L_e \ll l_p$ and the tube width $R
\ll L_e$ \cite{Odijk1983,Semenov1986}, both of which are functions of
$\phi$. The former is a measure of the average spacing between
adjacent collisions with background polymers, each of which contribute
an amount $k_B T$ to the average confinement energy of the test
chain. The latter measures the magnitude of the confined thermal
fluctuations (Fig.~\ref{fig:tube}).
Its mean value $\overline{R}$ as a function of monomer concentration
has been predicted theoretically based on a binary collision
approximation (BCA) and an effective medium approximation (EMA)
\cite{Morse2001}. The BCA focuses on the pairwise entanglement
topology of a test chain, while the EMA aims to account for the
collective network fluctuations. More recently, these mean-field type
theories have been challenged by the observation of pronounced
heterogeneities of the local tube width $R(s)$ along the tube contour,
which have been systematically studied in experiments
\cite{Kas1994,Dichtl1999,Romanowska2009,Wang2010,Glaser2010} and in
simulations \cite{Hinsch2007}.  The tube heterogeneities have been
statistically quantified by a broad and skewed tube width distribution
$P(R)$, which has been analyzed by an empirical model \cite{Wang2010}
and by a generalization of the BCA \cite{Glaser2010}.

In the following, we develop a systematic, BCA-based theory to
describe the fluctuations of the tube radius in an entangled polymer
solution on the scale of individual tube collisions. Thereby, the
local tube radius heterogeneities $R(s)$ and their distribution $P(R)$
can be determined, and $P(R)$ is found to be a universal non-Gaussian
scaling function with a stretched tail.  By comparison with the
segment fluid approach of Ref.~\cite{Glaser2010}, in which the
entangled solution is effectively mapped onto an ensemble of
entanglement segments, we predict the segment length $L$ (which was
previously treated as a fit parameter). The magnitude of the tube
width fluctuations is compared with published experimental data.  In
Ref.~\cite{Wang2010}, $P(R)$ was instead estimated based on an {\em
  ad-hoc} distribution of the local mesh size. The result turned out
to be unphysical at small values of $R$, however. As we show in the
following, the fluctuations of the tube radius $R$ can be
comprehensively described without additional assumptions based on a
generalization of the BCA.

The remainder of the paper is organized as follows. In
Sec.~\ref{sec:basic-elements}, we introduce the fundamental concept of
the tube and the wormlike chain (WLC) model for a single confined
semiflexible polymer.  In Sec.~\ref{sec:bca}, we then summarize the
basic assumptions underlying the BCA as an approximation to the
topological problem. Subsequently, in Sec.~\ref{sec:pphi}, we discuss
the statistical distribution of the confinement strength, which
explains the fluctuations of the tube radius, that are derived in
Sec.~\ref{sec:gaussianpr}.  In Sec.~\ref{sec:segment-fluid}, the
magnitude of tube radius fluctuations is used as an input to the
segment fluid model, which predicts a scaling function for the tube
radius distribution $P(R)$.  The analytical results are compared to
experimental data in Sec.~\ref{sec:exp}.

\section{Basic elements of the tube model}
\label{sec:basic-elements}

\subsection{Time scale separation and topology}
We consider a stiff test polymer in the presence of surrounding
uncrossable polymers, which are imposing topological constraints on
its conformation. We restrict our discussion to tightly entangled
polymers that are characterized by small transverse excursions around
an average path, the preferred contour.  The tube concept concerns
quantities in an intermediate equilibrium, i.e.\ on time scales
$\tau_e \ll t \ll \tau_d$, where $\tau_e$ is the time for the confined
degrees of freedom to equilibrate inside the tube and $\tau_d$ is the
disengagement time of the polymer from its initial tube.  In what
follows, a strong scale separation $\tau_e \ll \tau_d$ is assumed.
Then, in the idealized limit $\tau_d \to \infty$, the topological
relationships of the solution will be asymptotically conserved, and
the average positions of the background polymers and their mutual
topological relationships can be considered as effectively frozen (or
``quenched''), thus collectively giving rise to a (quasi-static)
confinement potential representing the tube. We denote the thermal
average with respect to a given ``quenched'' configuration by angular
brackets $\langle\cdots\rangle$ and the average over different
configurations and topologies of the tube by an over-bar
$\overline{\cdots}$. These ensemble averages correspond to temporal
averages over several time intervals of length $\tau_e$ and $\tau_d$,
respectively.

\subsection{Statistical mechanics of a single entangled stiff polymer}
\subsubsection{Transverse distance distribution}
\label{sec:transverse-distance}
To describe the physical properties of the test polymer, we
assume that the effect of confinement can, to leading order, be
described by a harmonic confinement potential. Hence, we use the
weakly-bending Hamiltonian
\begin{equation}
  H_{\mathrm{conf}} =
  \frac{l_p}{2} \int d s\,\left[\frac{\mathrm{d}^2 \vec r_\perp(s)}{\mathrm{d} s^2}\right]^2
  + \frac12 \int d s\,\phi(s)\vec r_\perp^2(s)
\label{eq:tube-hamiltonian}
\end{equation}
for the transverse fluctuations $\vec r_\perp(s)$ of the test polymer
about the straight ground state of a rigid rod, with a local
confinement strength $\phi(s)$ that will be determined
self-consistently. We use natural energy units ($k_B T=1$), such that
the persistence length $l_p=\kappa/k_B T$ is synonymous with the
bending rigidity $\kappa$.  We define the arc-length dependent tube
radius $R(s)$ via the variance of one component of the confined
transverse fluctuations,
\begin{equation}
R^2(s)\equiv \frac12 \langle \vec r_\perp^2(s) \rangle.
\label{eq:rsdef}
\end{equation}

Approximating the free energy by an effective Hamiltonian
$H_{\mathrm{conf}}$ that is quadratic in the fluctuations $\vec
r_\perp(s)$ is equivalent to approximating the distribution of $P[\vec
r_\perp(s)]$ in a given configuration by a Gaussian. Experiments
\cite{Dichtl1999,Glaser2010,Wang2010} and simulations
\cite{Zhou2006,Hinsch2007} indicate that the distribution of
transverse distances is indeed Gaussian for small transverse
displacements $\vec r_\perp(s)$ on the order of the tube radius $R$.
It can be shown theoretically, that this assumption is in accord with
a self-consistent treatment of the tube \cite{Morse2001}.

\subsubsection{Tube radius $R$ and entanglement length $L_e$}
\label{sec:rle}
As a first step, we consider the case of a test polymer in a
homogeneous (cylindrical) tube that can be characterized by the
spatial average $\overline{\phi}$ of a local confinement strength
$\phi(s)$, and hence a tube radius $R(s)\equiv R_0 = const\,$. The
variance of $\vec r_\perp(s)$ [Eq.~\eqref{eq:rsdef}] is obtained from
the tube Hamiltonian Eq.~\eqref{eq:tube-hamiltonian} via
equipartition, such that
\begin{eqnarray}
  \label{eq:rintmodes}
  R^2(s)&=& \int \frac{\dd q}{2 \pi} \frac{1}{l_p q^4 + \overline{\phi}}\\
  &=&\frac{1}{2 \sqrt{2} l_p^{1/4} \overline{\phi}^{3/4}} \equiv R_0^2
\label{eq:rint}
\end{eqnarray}
is the square of the tube radius corresponding to a homogeneous
confinement strength $\overline{\phi}$. Heterogeneities of the tube
potential and of the tube radius are discussed below (in Sections
\ref{sec:pphi} \& \ref{sec:gaussianpr}), where we show how small
spatial fluctuations $\delta\phi(s) \equiv \phi(s)-\overline{\phi}$
lead to spatial variations of $R(s)$ about its average value
$\overline{R}$.  We assume in what follows that the peak of the
corresponding distribution $P(R)$ is sufficently well defined such
that the average $\overline{R}$ and the typical value $R_0$ can be
used interchangeably.

The second characteristic quantity of the tube geometry, the
entanglement length $L_e$, is defined by assigning a harmonic
confinement energy equal to $k_B T$ ($=1$ in our units) to every
collision, and identifying $L_e$ with the collision length.  Writing
$\overline{\phi}$ for the average confinement potential strength in
Eq.~\eqref{eq:tube-hamiltonian}, equipartition yields
\begin{eqnarray}
  \label{eq:leint}L_e=\left[\int dq \frac{\overline{\phi}}{l_p q^4 + \overline{\phi}}\right]^{-1}=
  2\sqrt{2}\frac{l_p^{1/4}}{\overline{\phi}^{1/4}}.
\end{eqnarray}

\section{BCA}
\label{sec:bca}
We recapitulate the essential arguments that are needed to derive the BCA and
to understand the reasoning that follows.

The BCA was designed as an approximation to the underlying topological
many-body problem, suitable for estimating the absolute value of the
average tube radius $\overline{R}$, self-consistently. One considers
an elementary encounter (`binary collision') between two tubes,
calculates the free energy of confinement due to the uncrossability of
the chains, and sums over possible configurations of the pair of
tubes.  The key approximation consists in neglecting correlations
between multiple collisions.

\begin{figure}
\includegraphics[width=\columnwidth]{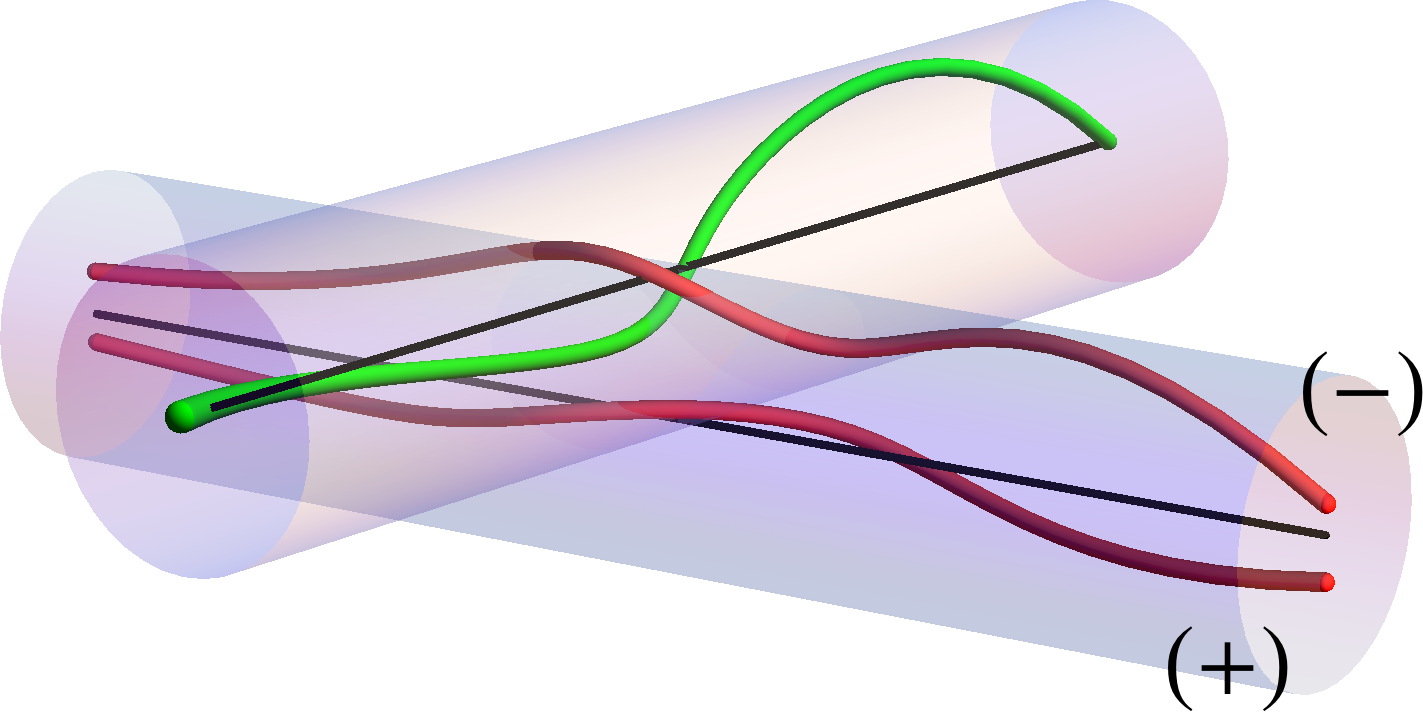}
\caption{(Color online) Topology of two confined polymers inside their
  tubes at a collision point: for fixed preferred tube contours in the
  transparent state, the polymers are found either in the `above'
  ($+$) or `below' ($-$) configuration. Adapted from
  Ref.~\cite{Morse2001}.}
\label{fig:top}
\end{figure}

The dynamic entanglement problem is cast into equilibrium statistical
mechanics language by assuming that the tube contours are
(temporarily) frozen. In such a configuration, an arbitrarily chosen
point on the tube contour associated with the test chain is
characterized by a Gaussian distribution of transverse distances, as
outlined in the preceding section. Its standard deviation is
approximated by the average tube radius $\overline{R}$. Consider now a
second background chain passing within a distance $\overline{R}$ from
the chosen point on the test chain. We refer to this event as a tube
collision. To calculate the contribution of the pair collision to the
confinement free energy, the BCA distinguishes between two states: a
hypothetical state, in which the test chain is transparent with
respect to collisions with the background chain and a state in which
the chains are mutually uncrossable. Due to the key assumption that
the collisions along the test chain are uncorrelated, the environment
of the collision point is completely random in the transparent state,
i.e.\ the distribution of transverse distances is unchanged from the
Gaussian distribution with standard deviation $\overline{R}$. The
average free energy of confinement follows from a topological
argument. In the uncrossable state for this pair of chains, the
configuration space of transverse fluctuations is divided into two
disconnected subspaces corresponding to the `above' ($+$) and the
`below' ($-$) configuration, as depicted in Fig.~\ref{fig:top}. The
average free energy in the uncrossable state is therefore obtained by
averaging the confinement potential in each of these subspaces over
the probability for a specific topology and collision geometry. The
resulting average free energy is equated with that of the transparent
state to obtain a self-consistent estimate of the tube radius
$\overline{R}$.

For the original mathematical implementation of the above ideas we
refer the reader to Ref.~\cite{Morse2001}. We note that a slightly
corrected estimate for the prefactor in the scaling result for the
average tube radius $\overline{R}$ with concentration was calculated
in Ref.~\cite{Glaser2010} (supplement).

\section{Distribution $P[\phi]$ of tube strengths}
\label{sec:pphi}
The original BCA is exclusively concerned with average values
$\overline{R}$, $\overline{\phi}$.  The scaling of these values with
concentration $\rho$ has been tested experimentally
\cite{Tassieri2008,Romanowska2009,Wang2010}, but the prefactor is
sensitive to the precise control of the experimental conditions and is
usually treated as a fit factor.  For a more detailed comparison of
theory, experiment and simulations, knowledge not only of the {\em
  average} value but of the richer and more robust tube radius {\em
  distribution} $P(R)$ is desirable (cf. Sec.~\ref{sec:gaussianpr}).

As a first step towards calculating $P(R)$, we derive the distribution
$P[\phi]$ of the local confinement strength $\phi(s)$ of a test
polymer.  Morse \cite{Morse2001} gave an explicit expression for the
confinement free energy of a test chain colliding with a medium
chain. We will explicitly adopt the mathematical approximation of
straight tube contours, that was implicitly made in the previous work,
and it is shown that inconsistencies resulting from this approximation
are avoided by verifying that the calculated quantities do not depend
on the overall chain length.  Let the distance of shortest approach
between two preferred contours with orientations $\vec u$, $\vec u'$
and centers of mass $\vec r$, $\vec r'$ be $x = (\vec r - \vec
r')\cdot \vec e_x$, where $\vec e_x = \vec u \times \vec u'/ |\vec u
\times \vec u'|$ is the direction perpendicular to both preferred
contours, then the free energy of the test polymer whose preferred
contour has been uniformly displaced by a vector $\vec h=h \vec e_h$ is
\begin{equation}
F_\pm(\vec h) = - \ln\Phi\left(\pm \frac{x- h \cos\psi}{\overline{R}}\right).
\label{eq:free-energy}
\end{equation}
Here, the sign $\pm$ refers to the specific topology
(cf. Fig.~\ref{fig:top}) and $\cos\psi=\vec e_h\cdot\vec e_x$.  The
function $\Phi(x)$ is given by the restricted partition sum of the
Gaussian fluctuations of the test polymer in the presence of an
uncrossable test chain \cite{Morse2001},
\begin{equation}
  \Phi(y) = \frac12 \mbox{erfc}\left(-\frac{y}{2}\right).
  \label{eq:defphi}
\end{equation}
It can be interpreted as the probability $p_\pm(x)=\Phi(\pm
x/\overline{R})$ of finding a specific topology.

From Eq.~\eqref{eq:free-energy}, we obtain the confinement strength
$\phi$ in a given configuration of preferred contours and topology as
the second derivative of the free energy,
\begin{equation}
  \phi_\pm(x) = \frac{\mathrm{d^2}}{\mathrm{d} h^2} F_\pm(\vec h)\Big|_{h=0}
  =-\frac{\cos^2\psi}{\overline{R}^2} \frac{\mathrm{d}^2}{\mathrm{d} y^2}\ln \Phi(y) \Big|_{\pm x/{\overline{R}}}.
\end{equation}

\begin{figure}
\centering\includegraphics[width=\columnwidth]{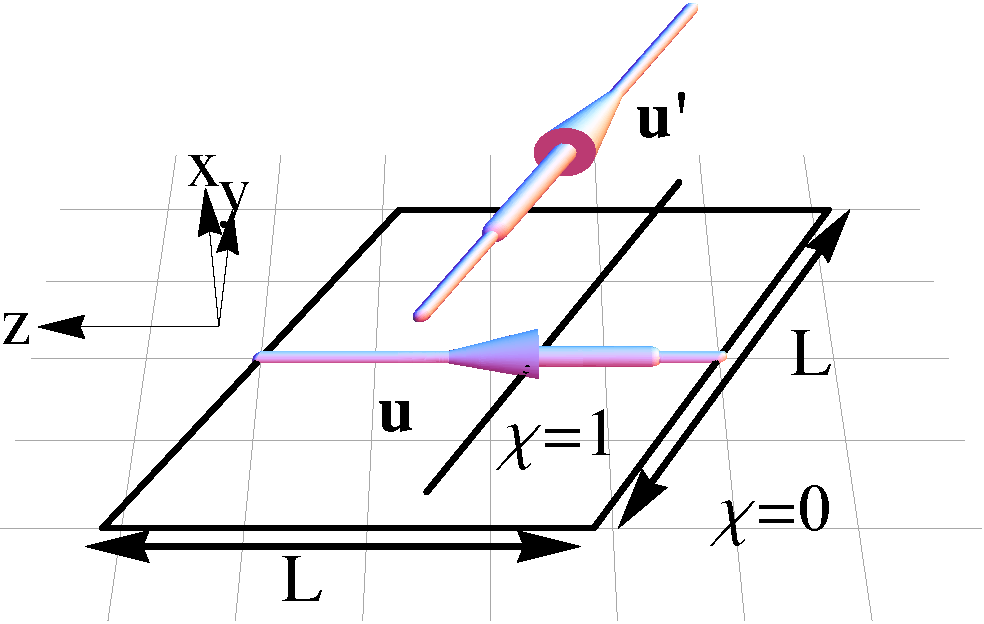}
\caption{(Color online) Overlap area (parallelogram) of two tubes of
  length $L$, represented by their preferred contours with
  orientations $\vec u$, $\vec u'$ enclosing an arbitrary angle, and
  characteristic function $\chi$. }
\label{fig:overlap}
\end{figure}

To derive the distribution of confinement strengths $\phi(s)$, we now
turn back to the central BCA approximation that the collisions between
the test polymer and the background polymers are independent localized
events.  Due to the requirement $\overline{R}\ll L_e$, we may, without
loss of generality, even treat them as point-like and express the
confinement potential $\phi(s)$ per unit length at a point $s$ on the
test polymer as
\begin{equation}
  \phi(s) = \sum_{i=1}^N \chi(\vec r_i, \vec u_i,\vec u) \delta (s -z_i) \phi_\pm (x_i).
\label{eq:philocal}
\end{equation}
Here, $\chi(\vec r_i,\vec u_i,\vec u)$ is a characteristic function of overlap
between the colliding tubes, which takes on the value one whenever a
tube collision occurs, and zero otherwise (for a graphical definition
see Fig.~\ref{fig:overlap}). We introduce it, here, merely as a
convenient tool to facilitate the formal manipulation of the
following expressions. The coordinate $z_i$ in the argument of the
$\delta$-function is the point of shortest approach (the collision
point) between the two tubes on the test polymer.

The distribution $P[\phi]$ of confinement strengths that follows
from Eq.~\eqref{eq:philocal} is of the Holtsmark type
\cite{Holtsmark1919,Simon1990} and describes the total confinement
potential of the test chain as a sum of contributions resulting from
uncorrelated collisions with medium chains.  Explicitly, its
moment-generating functional is given by (cf. App.~\ref{app:pphi})
\begin{equation}
  P[w(s)] = \exp\left[\frac{n}{2 \pi}\sum_\pm \int\vec d \vec r'\int'\! d\vec u'\,\chi e^{i w(z') \phi_\pm(x')} p_\pm(x')\right].
\label{eq:pphi-characteristic}
\end{equation}
From this, we obtain the average of $\phi(s)$ by functional
differentiation of the logarithm of Eq.~\eqref{eq:pphi-characteristic}
with respect to the field $w(s)$ (cf. App.~\ref{app:pphi}),
\begin{equation}
  \overline{\phi}=\frac{nL \pi}{8 \overline{R}} \int d \tilde x\,\frac{\Phi'^2(\tilde
    x)}{\Phi(\tilde x)}.
\label{eq:phibar}
\end{equation}
The integral in Eq.~\eqref{eq:phibar} is numerically evaluated and
gives $\overline{\phi}=\alpha \, nL/\overline{R}$, with
$\alpha=0.502$, in agreement with earlier results \cite{Morse2001}
(applying the slight numerical correction discussed in
Ref.~\cite{Glaser2010}, supplement).

We proceed analogously to obtain the second cumulant
(cf. App.~\ref{app:pphi}), the correlation function
\begin{equation}
  \overline{\phi(s)\phi(s')} = \beta \frac{nL}{\overline{R}^3} \delta(s-s'),
\label{eq:phiphi}
\end{equation}
where $\beta = 0.0941$. The uncorrelated character of the collisions
is apparent from the $\delta$-function on the RHS of
Eq.~\eqref{eq:phiphi}. Both should be understood in the coarse-grained
sense, assuming $\overline{\phi(s)\phi(s')}=0$ when $s-s' \gg
\overline{R}$.

\section{Gaussian approximation to the tube radius distribution $P(R)$}
\label{sec:gaussianpr}
We can now turn the distribution $P[\phi]$, which we characterized by
its first two cumulants, into a Gaussian approximation to the tube
radius distribution $P(R)$, by calculating the linear response of the
local tube radius $R(s)$ to spatial changes (heterogeneities) in
$\phi(s)$. We begin with the observation that the correlation function
of the (projected) transverse fluctuations, $C(s,s') = \langle \vec
r_\perp(s) \cdot \vec r_\perp(s')\rangle/2$ obeys the following
differential equation (cf. App.~\ref{app:heterogeneities})
\begin{equation}
-l_p \partial_s^4 C(s,s') - \phi(s') C(s,s') = \delta(s-s')
\label{eq:correlations}
\end{equation}
The tube radius is given by $R(s)=[C(s,s)]^{1/2}$ and the linear response
expression for this quantity is calculated in
App.~\ref{app:heterogeneities} as
\begin{equation}
  R(s) = \overline{R} - \frac{1}{2 \overline{R}} \int ds'\, G^2(s-s') \delta \phi(s').
\label{eq:linearresponse}
\end{equation}

The variance of the tube radius $R(s)$ at a randomly chosen point $s$
on the test polymer is now calculated from
Eq.~\eqref{eq:linearresponse} as
\begin{equation}
\begin{aligned}
  &\overline{\delta R^2}\equiv \overline{[R(s) - \overline{R}]^2}\\
  &= \frac{1}{4 \overline{R}^2} \int ds' ds''\, G^2(s-s') G^2(s-s'')
  \overline{\phi(s') \phi(s'')}.
\end{aligned}
\label{eq:variance}
\end{equation} 
Within the BCA, with its trivial spatial correlations of the
confinement potential $\phi(s)$, Eq.~\eqref{eq:phiphi}, this reduces to
\begin{equation}
\label{eq:g4}
\overline{\delta R^2} = \beta \frac{nL}{4 \overline{R}^5} \int ds' G^4(s-s').
\end{equation}
The integral Eq.~\eqref{eq:g4} is numerically evaluated using an explicit
expression for $G(s)$ [Eq.~\eqref{eq:gexplicit}],
\begin{equation}
\int ds\, G^4(s) = \gamma \, \overline{R}^8 L_e,
\label{eq:intg4}
\end{equation}
where $\gamma=0.5125$, which yields the final result for the variance of $P(R)$,
\begin{equation}
\overline{(\delta R)^2}=\frac14 \beta \gamma nL \overline{R}^3 L_e.
\label{eq:variancefinal}
\end{equation}

Using the self-consistent solutions for $\overline{R} = (4
\alpha)^{-3/5} (nL)^{-3/5} l_p^{-1/5}$ and $L_e = (\alpha/8)^{-2/5}
(nL)^{-2/5} l_p^{1/5}$ \cite{Morse2001}, mean and variance of the
Gaussian approximation to the tube radius distribution $P(R)$ are
completely determined in terms of the contour length concentration
$nL$ and the persistence length $l_p$. In particular, the coefficient
of variation $c_v=[\overline{(\delta R)^2}]^{1/2}/\overline{R}$ turns
out to be a concentration-independent constant,
\begin{equation}
c_v = \frac12 \sqrt{\frac{\beta \gamma}{\alpha}} =  0.155.
\label{eq:cov}
\end{equation}

\section{Segment fluid approximation}
\label{sec:segment-fluid}
In Ref.~\cite{Glaser2010}, a broad distribution $P(R)$ of the tube
radius was found. The derivation of an analytical result for this
distribution was also based on a Holtsmark-type distribution for the
confinement strength $\phi$ resulting from uncorrelated collisions,
but the latter were averaged over the characteristic length $L$ of
entanglement segments (which is why the approach was called a
``segment-fluid'' approximation). It was argued that this length is on
the order of $L_e$. We now show that this choice is indeed justified
and predict the precise value of the segment length.

\begin{figure}
\centering\includegraphics[width=\columnwidth]{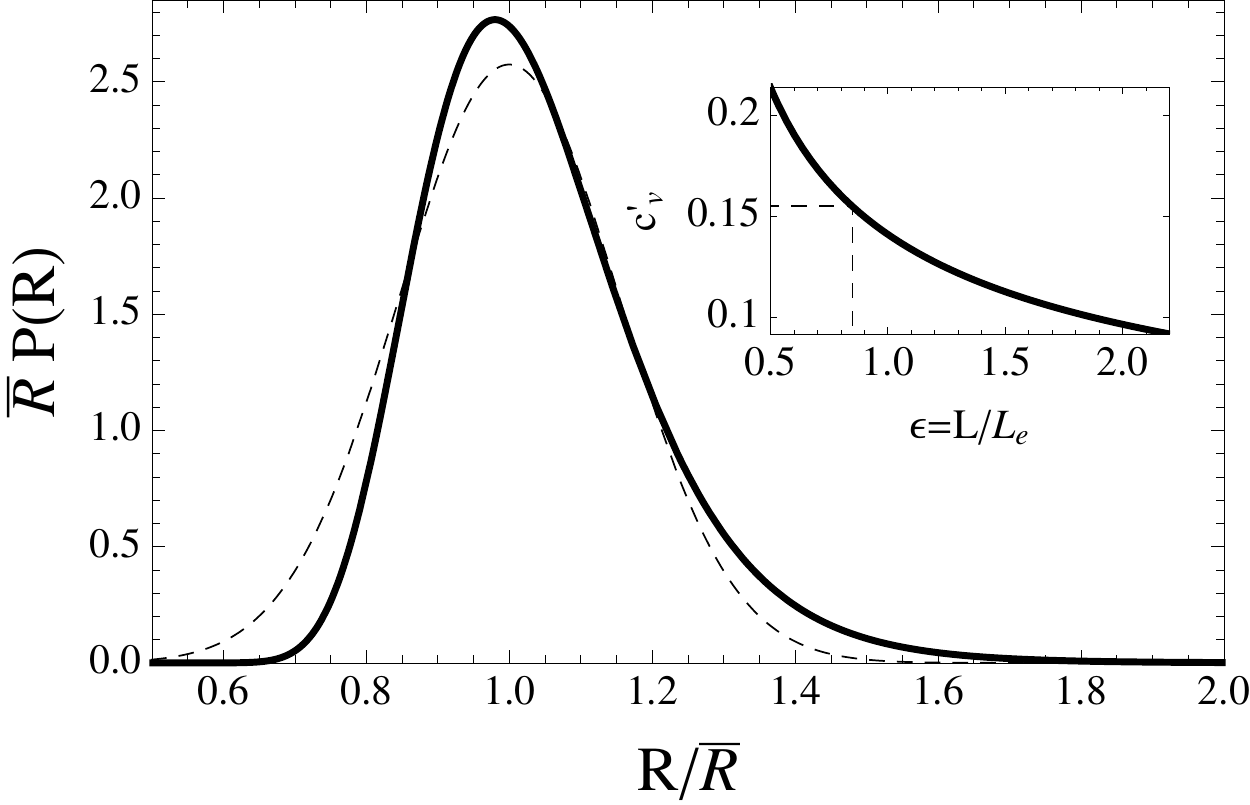}
\caption{(Color online) Reduced distribution $\overline{R}
  P(R/\overline{R})$ of the tube radius $R$. The solid line is the
  segment-fluid prediction, the dashed line is a Gaussian
  approximation with coefficient of variation $c_v$ given by
  Eq.~\eqref{eq:cov}. {\em Inset:} Coefficient of variation $c'_v$ of
  the segment fluid approximation versus reduced segment length
  $\epsilon=L/L_e$, and the predicted value.}
\label{fig:pr}
\end{figure}

The tube radius distribution $P(R)$ was given as an analytical
approximation in Ref. \cite{Glaser2010},
\begin{equation}
P(R)=\frac{8}{3 R \Gamma(k)} \exp(-y)\,y^k,\quad y\equiv
0.01325\, \frac{L\overline{R}^2}{\ell_p^{1/3} R^{8/3}},
\label{eq:pr}
\end{equation}
where $k = 4.013 n' L^2 \overline{R}$, $n'$ is the number density of
entanglement segments and $\Gamma(x)$ is the Gamma function.  It was
noted \cite{Glaser2010} that $P(R)$ can be written as a scaling
function $P(R)=(1/\overline{R}) f(R/\overline{R},n' L^2
\overline{R})$.  This implies that the coefficient of variation is
given by a constant, $c'_v = g(n' L^2 \overline{R})$.  If we make the
ansatz $L=\epsilon L_e$, with an undetermined dimensionless constant
$\epsilon$ and use the self-consistent values for $\overline{R}$ and
$L_e$ given above, we get $c'_v =g(\epsilon/\alpha)$.  The function
$g(x)$ is easily evaluated numerically and is shown in
Fig.~\ref{fig:pr} (inset). To fix $\epsilon$ and thus the length $L$
of entanglement segments, we require $c'_v = c_v$ and obtain
numerically $\epsilon=0.85$.

Evaluating the tube radius distribution $P(R)$ [Eq.~\eqref{eq:pr}] at
this value of the reduced segment length $L/L_e$, we obtain $k=y
(R/\overline{R})^{8/3}=6.79$ and hence all parameters occurring in
$P(R)$ are now fully specified, giving
\begin{equation}
  P(R) =  N \exp\left[-6.79 \left(\frac{\overline{R}}{R}\right)^{8/3}\right]
(\overline{R}/R)^{8/3},
\label{eq:prfinal}
\end{equation}
with a normalization constant $N=2.434\times 10^3$. The corresponding
unique reduced distribution $\overline{R} P(R)$ is shown in
Fig.~\ref{fig:pr} and compared to the Gaussian distribution with the
same $c_v$. Beyond what was achieved in Ref.~\cite{Glaser2010}, the
functional form of $P(R)$ is now fully determined.  As can be seen in
Fig.~\ref{fig:pr}, the distribution $P(R)$ is positively skewed and
has a broad tail at large values of $R$.

\section{Comparison to experiment}
\label{sec:exp}
\begin{figure}
\centering\includegraphics[width=\columnwidth]{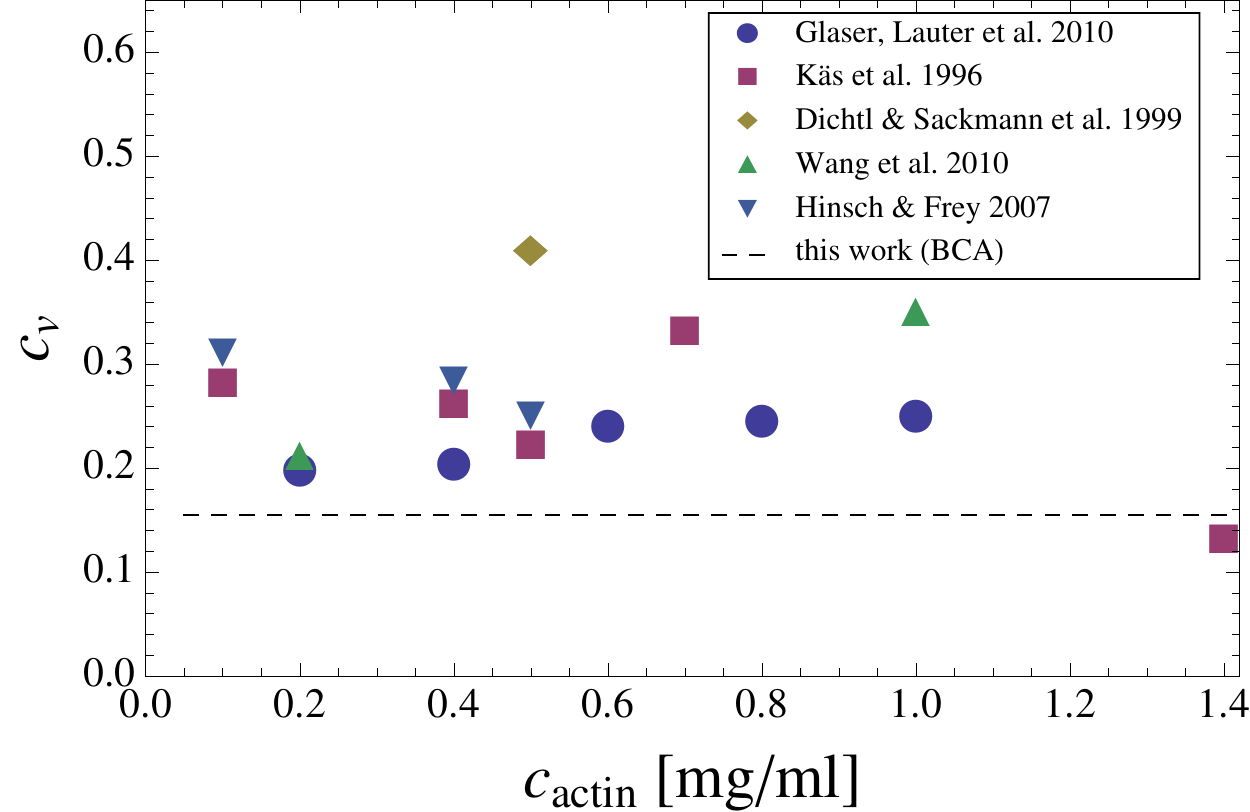}
\caption{(Color online) Comparison of tube radius fluctuations in different
  experiments and simulations of semidilute solutions of
  F-actin. Shown is the coefficient of variation
  $c_v=\sqrt{\overline{\delta R^2}}/\overline{R}$ for the
  fluctuations of $R$. Symbols correspond to data taken from the
  literature. Circles: experimental data taken from
  Ref.~\cite{Glaser2010} (Fig.~3, inset); squares: experimental data
  from Ref.~\cite{Kas1996}; diamonds: experimental data from
  Ref.~\cite{Dichtl1999} for the fluctuations of the response
  coefficient $\alpha_\perp$, converted to fluctuations of the tube
  radius $R$; upright triangles: experimental data from
  \cite{Wang2010} (Fig.~2); downward facing triangles: simulation data
  from Ref.~\cite{Hinsch2007} (Fig.~9). The solid dashed line is the
  prediction of Eq.~\eqref{eq:cov}.}
\label{fig:cov}
\end{figure}

The functional form of $P(R)$ was compared to experimental data in
Ref.~\cite{Glaser2010}, and very good qualitative agreement was found,
using the value of the segment length $L$ as a fit
parameter. Remarkably, also our above prediction of a constant value
for the coefficient of variation $c_v$, which can
be checked against a whole set of independent measurements, is
nicely confirmed by the data (Fig.~\ref{fig:cov}).

The figure summarizes literature data for $c_v$ against monomer
concentration $c$ from various experiments and simulations of
semidilute actin solutions. The dashed line is our prediction from
Eq.~\eqref{eq:cov}. Two results are evident from this plot. First, the
data scatter within a band of $c_v=0.2$-$0.4$. Second, the
theoretical prediction lies below most of the data points and thus
provides a lower bound for the observed tube radius fluctuations. This
suggests that a constant value for the coefficient of variation is
indeed consistent with the reported data, but that the heterogeneities
are actually about twice as strong as predicted.

This is not entirely unexpected, since the BCA, on which our
theoretical derivation relies, is not meant to describe the {\em
  absolute} value of the tube radius quantitatively. In fact, the BCA
alone is well known to underestimate the tube fluctuations, since it
does not take into account the collective fluctuations of the
surrounding medium into which the tube is embedded
\cite{Morse2001}. Corresponding quantitative discrepancies with
experiments have been reported before \cite{Tassieri2008}.

\section{Conclusion}
We have calculated the fluctuations of the tube radius in entangled
solutions of semiflexible polymers, based on the binary collision
approximation (BCA). We predict that the shape of the tube radius
distribution is given by a universal (concentration-independent)
scaling function, for which we gave an analytical approximation in
Eq.~\eqref{eq:prfinal}. Our results provide a quantitative
characterization of the local packing structure of entangled biopolymer
solutions in terms of distribution functions, which are at the same
time a more sensitive and more robust means for comparing data and
theory than average values alone.  We hope that the methods of
analysis established here may find application in future experimental
studies, e.g.\ in microrheology \cite{Waigh2005,Wirtz2009}, or in the
interpretation of simulation data \cite{Hinsch2007,
  Ramanathan2007}. Further theoretical questions, such as the
characterization of the distribution of tube contours
\cite{Hinsch2009,Romanowska2009}, are currently under study.

\begin{acknowledgments}
  We are indebted to David Morse for his most valuable comments on the
  manuscript.  We also gratefully acknowledge inspiring discussions
  with Lars Wolff, Dipanjan Chakraborty, Sebastian Sturm and Andrew
  Gustafson.  We thank Inka Lauter for providing experimental data on
  the tube width fluctuations of actin solutions that motivated the
  discussion of the tube radius distribution.  This work was supported
  by the Deutsche Forschungsgemeinschaft (DFG) through FOR 877 and the
  Leipzig School of Natural Sciences -- Building with Molecules and
  Nano-objects.
\end{acknowledgments}

\begin{appendix}

\section{Calculation of the distribution of confinement strengths
$P[\phi]$}
\label{app:pphi}
The distribution of the local confinement strength is formally obtained as
the average $P[\varphi(s)] =
\overline{\delta[\phi(s)-\varphi(s)]}$ over all possible
configurations of tube contours and topologies.  The corresponding
characteristic functional, $P[w(s)]$, follows from a functional
Fourier transform and Eq.~\eqref{eq:philocal} as
\begin{equation}
P[w(s)] = \overline{\exp\left(i \sum_{i=1}^N \chi(\vec r_i, \vec u_i, 
\vec u) w(z_i) \phi_\pm (x_i)\right)}.
\end{equation}
Here, tangents to the test and the background chains' tube backbones
are denoted by $\vec u$ and $\vec u_i$, and the vector $\vec r_i$
connects the tubes' center of mass of the test chain with that of the
background chain $i$. Its coordinates $x_i$, $y_i$ and $z_i$ are
defined along the directions $\vec e_{x,i}=(\vec u\times \vec
u_i)/|\vec u\times \vec u_i|$, $\vec e_{y,i}=\vec u\times \vec
e_{x,i}$ and $\vec e_{z,i} = \vec u$.  The quenched average
$\overline{\cdots}$ is implemented as the simultaneous average over
the probability $p_{\pm}(x_i)=\Phi(\pm x_i)$ of finding a specific
topology `+' or '-' [defined after Eq.~\eqref{eq:defphi}] and over the
uniformly distributed centers of mass and orientations of the
tubes. Since the chains' preferred contours are assumed to be
uncorrelated, the average over the $N$ preferred contours and the
topology of the background chains relative to the test chain
factorizes as
\begin{equation}
P[w(s)]=\left[\sum_\pm \int \frac{d\vec r'}{V} \int'\frac{d\vec u'}{2 \pi}
e^{i \chi(\vec r', \vec u',\vec u) w(z') \phi_\pm(x')}p_\pm(x')\right]^N.
\label{eq:uncorrelated}
\end{equation}
Here, the integral over orientations extends over the half-sphere
(indicated by a prime). Exploiting the formal definition of $\chi$ as
a characteristic function of overlap, which amounts to setting the
factor in the brackets for non-overlapping chains to unity (since the
probability $p_\pm(x')$ is normalized), we get
\begin{equation}
\begin{aligned}
  &P[w(s)]\\
  &=\left\{1+\sum_\pm \int \frac{d\vec r'}{V} \int'\frac{d\vec u'}{2
      \pi} \left[e^{i w(z')
        \phi_\pm(x')}p_\pm(x')-1\right]\chi\right\}^N.
\end{aligned}
\end{equation}
Using $n=N/V$ for the polymer number concentration and performing the
limit $N\to\infty$, Eq.~\eqref{eq:pphi-characteristic} in the main
text is obtained.

The first cumulant is obtained by functional differentiation of the
characteristic functional Eq.~\eqref{eq:pphi-characteristic} with
respect to the field $w(s)$,
\begin{align}
  \overline{\phi(s)} &=- i \frac{\delta}{\delta w(s)} \ln P[w(s)]\Big|_{w(s)=0}\\
  &=\frac{n L}{2 \pi} \sum_{\pm} \int' d\vec u'\, \int d \vec
  r'\,\delta(z'-s)\chi p_\pm(x')\phi_\pm(x').
\label{eq:1stcumulant}
\end{align}
Using the fact that the integral of $\chi\delta(z'-s)$ over $dy'$ and
$dz'$ is the height $L \sin \theta$ of the overlap area
(Fig.~\ref{fig:overlap}), carrying out the second derivative of
$-\ln\Phi$ and the angular integrals, and using $\sum_\pm
\Phi(x') =1$, one arrives at Eq.~\eqref{eq:phibar} in the main
text. Analogously, we obtain the second cumulant, the correlation
function
\begin{align}
&  \overline{\phi(s)\phi(s')}\\
& = - \frac{\delta^2}{\delta w(s) \delta w(s')}
\ln P[w(s)]\Big|_{w(s)=0}\\
&= \frac{n L}{2 \pi}
  \sum_{\pm} \int' d\vec u'\, \int d \vec r'\,\delta(z'-s)\delta(z'-s')\chi
 p_\pm(x')\phi_\pm^2(x').
\end{align}
Applying a similar reasoning as above to simplify the equation
and numerically evaluating the remaining $x$-integral, we obtain
Eq.~\eqref{eq:phiphi} in the main text.

\section{Heterogeneous tube radius $R(s)$}
\label{app:heterogeneities}
The fluctuation-response relation Eq.~\eqref{eq:correlations} for the
correlation function $C(s,s')$ is derived from the free energy $- \ln
Z[\vec f_\perp(s)]$ of a confined WLC in the presence of an external
transverse force $\vec f_\perp(s)$. The corresponding Hamiltonian is
$H=H_{\mathrm{conf}}+ \int ds\,\vec f_\perp(s) \vec r_\perp(s)$. Since
\begin{align}
\langle \vec r_\perp(s) \rangle &= \frac{\delta \ln Z}{\delta \vec f_\perp(s)},\\
\langle \vec r_\perp(s) \vec r_\perp(s') \rangle &= \frac{\delta \ln Z}{\delta
\vec f_\perp(s)\delta \vec f_\perp(s')}\Big|_{\vec f_\perp(s)=0}\\
&=\frac{\delta \langle \vec r_\perp(s) \rangle}{\delta \vec f_\perp(s')}\Big|_{
\vec f_\perp(s)=0},
\end{align}
it follows that $\vec C(s,s')\equiv \langle \vec r_\perp(s) \vec
r_\perp(s') \rangle$ is the functional inverse of $\vec
C(s,s')^{-1}=\delta \vec f_\perp(s)/\delta {\langle} \vec r_\perp(s')
{\rangle}$. Since the force $ \vec f_\perp(s) = -{\langle} \delta
H_{\mathrm{conf}}/\delta \vec r_\perp(s) {\rangle}$ producing an
average displacement ${\langle}\vec r_\perp(s){\rangle}$ is given by
$-l_p \partial_s^4 \langle \vec r_\perp(s) \rangle - \phi(s) \langle
\vec r_\perp(s) \rangle$, Eq.~\eqref{eq:correlations} follows by
partial integration.

A solution of Eq.~\eqref{eq:correlations} would exactly describe the
tube heterogeneities that follow from a heterogeneous confinement
potential $\phi(s)$. However, no such solution is available for
arbitrary $\phi(s)$. Therefore, we write $\phi(s) = \overline{\phi}
+\delta \phi(s)$ with small fluctuations $\delta\phi(s)$ about the
average confinement strength $\overline{\phi}$. A simple
first-order perturbation scheme for $C(s,s') = C^{(0)}(s,s') + \delta
C(s,s')$ is set up by requiring $C^{(0)}$ to be the response function
in the homogeneous case, where $\delta\phi(s)=0$,
\begin{equation}
  C^{(0)}(s-s') = G(s-s')= \int\frac{dq}{2 \pi} \frac{e^{i q (s-s')}}{l_p q^4 + \overline\phi}.
\label{eq:fourier}
\end{equation}
The explicit expression for the Fourier transform in
Eq.~\eqref{eq:fourier} can be obtained analytically and written [using
Eqs.~\eqref{eq:rint} \&~\eqref{eq:leint}] as
\begin{equation}
G(s-s') = R_0^2\, e^{-2 |s|/L_e} \left[\cos\left(2\frac{s}{L_e}\right)
 + \sin\left(2 \frac{|s|}{L_e}\right)\right].
\label{eq:gexplicit}
\end{equation}

The leading-order response $\delta C(s,s')$ to the perturbation
$\delta\phi(s,s')$ is obtained from Eq.~\eqref{eq:correlations} if
small terms $\mathcal O(\delta C\delta\phi)$ are neglected,
\begin{equation}
  \delta C(s,s') = - \int ds'' G(s-s'') \delta \phi(s'') G(s'-s'').
\label{eq:deltac}
\end{equation}
Eq.~\eqref{eq:linearresponse} in the main text is obtained by writing
$R(s) = \overline{R} + \delta R(s)$ with $\delta R(s)=\delta(R^2)/2
\overline{R} = \delta C(s,s)/2 R_0$.

\end{appendix}

\end{document}